\documentclass[aps, pra, superscriptaddress,preprintnumbers,twocolumn]{revtex4-2}
\usepackage{graphicx}
\usepackage{dcolumn}
\usepackage{bm}
\usepackage{amsmath}
\usepackage{amsfonts}
\usepackage[svgnames]{xcolor}
\usepackage{appendix}
\usepackage{multirow}
\usepackage{booktabs}
\usepackage{tabularx}
\usepackage{amsthm}
\usepackage{setspace}
\usepackage[T1]{fontenc}
\usepackage{extarrows}
\usepackage{lineno}
\usepackage[percent]{overpic}
\usepackage{dsfont}
\usepackage{physics}
\graphicspath{{fig_export/}}
\DeclareGraphicsExtensions{.pdf,.png} 
\usepackage[colorlinks,
            linkcolor=blue,
			filecolor=blue,
			urlcolor= blue,
			citecolor=blue,
            ]{hyperref}
\usepackage{algpseudocode}
\usepackage{orcidlink}

\newcounter{algorithm}
\renewcommand{\thealgorithm}{\arabic{algorithm}}

\makeatletter
\newcommand{\pralgmakecaption}[1]{%
    \par\vspace{6pt}%
    \begingroup
    \@parboxrestore
    \small
    \@makecaption{\textbf{Algorithm \thealgorithm}}{#1}\par
    \endgroup
}
\makeatother

\algrenewcommand\algorithmicrequire{\textbf{Input:}}
\algrenewcommand\algorithmicensure{\textbf{Output:}}
\definecolor{revisionred}{RGB}{177,39,71}

\begin{document}

\title{Pontryagin's principle for leakage-immune adiabatic quantum state transfer}

\author{Xiao-Yu Dong\,\orcidlink{0009-0006-4937-3539}}
\affiliation{State Key Laboratory of Semiconductor Physics and Chip Technologies, Institute of Semiconductors, Chinese Academy of Sciences, Beijing 100083, China}
\affiliation{Center of Materials Science and Opto-Electronic Technology, University of Chinese Academy of Sciences, Beijing 100049, China}

\author{Xi-Lai Wang}
\affiliation{State Key Laboratory of Semiconductor Physics and Chip Technologies, Institute of Semiconductors, Chinese Academy of Sciences, Beijing 100083, China}
\affiliation{Center of Materials Science and Opto-Electronic Technology, University of Chinese Academy of Sciences, Beijing 100049, China}

\author{Wen-Long Ma\,\orcidlink{0000-0002-5080-407X}}
\email{wenlongma@semi.ac.cn}
\affiliation{State Key Laboratory of Semiconductor Physics and Chip Technologies, Institute of Semiconductors, Chinese Academy of Sciences, Beijing 100083, China}
\affiliation{Center of Materials Science and Opto-Electronic Technology, University of Chinese Academy of Sciences, Beijing 100049, China}

\date{\today}

\begin{abstract}
The standard stimulated Raman adiabatic passage (STIRAP) protocol enables high-fidelity quantum state transfer in an ideal three-level system via adiabatic following of a dark state. However, in practical systems with more energy levels, control pulses with finite spectral selectivity often couple the three-level subspace to the remaining subspace, introducing leakage that fundamentally limits the transfer performance. Here, we adopt a multilevel chain model for STIRAP that explicitly incorporates this leakage subspace. Using Pontryagin's maximum principle, we formulate a leakage-penalized quantum optimal control problem with the control pulses constrained to experimentally feasible Gaussian pulse families. We derive explicit gradients of the objective functional with respect to the pulse parameters, enabling efficient low-dimensional optimization that suppresses leakage while preserving the counterintuitive STIRAP pulse ordering. Numerical simulations for a superconducting transmon platform demonstrate that the optimized control pulses significantly enhance the target-state transfer fidelity and improve robustness against amplitude miscalibration and detuning drifts.
\end{abstract}

\maketitle

% ========= main sections =========
\section{Introduction}
\label{Sec:intro}

Scalable quantum technologies require accurate and robust quantum state transfer \cite{Glaser2015TrainingSchrodingersCat, Mahesh2023QuantumOptimalControl}.
A central example is stimulated Raman adiabatic passage (STIRAP), where population is transferred through adiabatic following of a dark state while the occupation of the intermediate state is suppressed \cite{Bergmann1998,Vitanov2001,Bergmann2015,Vitanov2017}.
In realistic multilevel systems, however, the ideal three-level picture is often incomplete.
Finite drive amplitudes, limited spectral selectivity, and nearby leakage levels can distort the dark-state pathway and reduce the transfer fidelity \cite{Werninghaus2021npjQILeakageOptimalControl,Hyyppa2024PRXQuantumLeakageAnalyticalEnvelopes,Fischer2023PRXQuantumQuditGateSynthesis,PhysRevApplied.19.064071}.
Control performance is further degraded by fluctuations, dissipation, and unwanted transitions \cite{Yatsenko2014,Hou2013,Blekos2020,PRXQuantum.4.030312}.
These effects are especially relevant in weakly anharmonic superconducting circuits, where the same control fields that drive the target transitions may also couple to higher unwanted levels \cite{Koch2007,Blais2021,Lacroix2025PRLFastFluxActivatedLRU}.
The problem addressed in this work is therefore how to preserve the robustness of STIRAP while optimizing experimentally constrained control pulses to suppress leakage.

\begin{figure}[t]
\centering
\includegraphics[width=\linewidth]{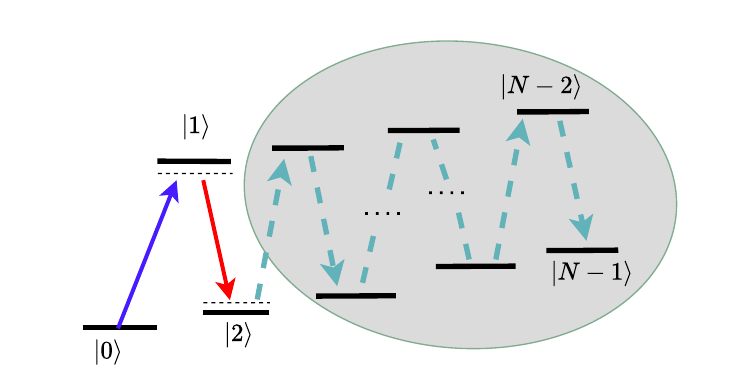}
\caption{Multilevel chain model for STIRAP-type population transfer with leakage.
The desired transfer pathway consists of states \(|0\rangle\), \(|1\rangle\), and \(|2\rangle\), while the shaded manifold represents higher-lying leakage states.
The blue and red arrows denote the two intended couplings along the transfer pathway, and the dashed arrows indicate successive off-target couplings responsible for leakage.}
\label{fig:chain_model}
\end{figure}

Pontryagin's maximum principle (PMP) provides a useful framework for constrained quantum-control problems \cite{Liberzon2012COVandOCT,PRXQuantum.2.030203}.
It treats the optimization as a two-point boundary-value problem.
In this formulation, the quantum state is propagated forward, while a costate is propagated backward
\cite{Ansel_2024}.
The resulting optimality conditions give control gradients for numerical pulse updates.
Related gradient-based methods, such as gradient ascent pulse engineering (GRAPE) \cite{deFouquieres2011SecondOrderGRAPE,Khaneja2005GRAPE} and Krotov-type methods \cite{Reich2012KrotovJCP}, have shown the power of gradient information in quantum control
\cite{Peirce1988PRAOCT,Palao2002UnitaryOCT,Brif2010ControlPerspective,dalessandro2021introduction,PhysRevA.77.063420}.
PMP provides a unified framework in which the dynamics, running costs, and control constraints can be incorporated on the same footing \cite{Li2024PRAOptimalSTIRAP}.
This makes PMP suitable for multilevel STIRAP with experimental pulse constraints.
This motivates a PMP-based optimization of a small set of pulse parameters to suppress leakage while preserving robust state transfer.
The multilevel chain structure considered in this work is illustrated in Fig.~\ref{fig:chain_model}.

Chain-based STIRAP models provide a natural way to describe dark-state transfer in multilevel systems.
In the ideal three-level case, STIRAP transfers population between two endpoint states by adiabatically following a dark state, while the intermediate state remains weakly occupied
\cite{Kuklinski1989,Gaubatz1990,Kral2007}.
Multilevel dark-state transport and chain-based generalizations provide the natural framework for extending this picture beyond an isolated three-level manifold \cite{PembertonRoss2010DarkStates,Kumar2012DarkStateManifold}.
This dark-state mechanism has motivated optimized STIRAP-type schemes in few-level molecular settings \cite{PhysRevA.86.023406,PhysRevA.80.042325}, as well as superconducting implementations of STIRAP-type population transfer \cite{Kumar2016,Xu2016}.
Superadiabatic passage and related optimal-control or STIRAP-inspired protocols have further extended these ideas in superconducting multilevel platforms \cite{Vepsalainen2019,Zheng2022npjQI_OCT_STIRAP_Qudit,Niu2022PRAppSTIRUP,Singhal2025PRAppSTIRAPInspiredGatesDualRail}.
However, realistic systems rarely form an isolated three-level manifold.
In weakly anharmonic superconducting circuits, finite drive amplitudes can couple the target manifold to nearby leakage levels, and ac-Stark shifts or limited spectral selectivity can further deform the dark-state pathway \cite{Goerz2017Charting}.
Analytic pulse-shaping methods, such as the derivative removal by adiabatic gate (DRAG), can reduce leakage in weakly anharmonic circuits~\cite{Motzoi2009,Gambetta2011}, while shortcut-to-adiabaticity methods can suppress nonadiabatic errors and accelerate transfer \cite{Demirplak2008,Berry2009,GueryOdelin2019}.
These methods are powerful, but additional fields or waveform complexity may introduce unwanted transitions and calibration sensitivity in multilevel devices \cite{Li2016,Du2016,Evangelakos2023STIRAPShortcuts,Baksic2016,Cepaite2023COLD}.
PMP offers a complementary route because it can incorporate leakage penalties, running costs, and pulse constraints within the same optimization framework.
It is therefore natural to combine a chain-based STIRAP description with PMP-based parameter optimization, so that experimentally feasible pulses can be adjusted to suppress leakage while retaining the robustness of dark-state transfer.

In this work, we address how STIRAP-type state transfer can be optimized in a realistic multilevel system where leakage and experimental pulse constraints are present.
To address this problem, we describe the dynamics in a rotating frame and under the rotating-wave approximation using an $N$-state nearest-neighbor chain model.
This model includes both the target transfer pathway and the dominant leakage pathways in the same Hamiltonian.
We then cast the pulse-design task as a leakage-penalized Bolza optimal-control problem and derive the PMP state-costate equations and control gradients.
The control fields are restricted to Gaussian pump and Stokes pulses, so the optimization is reduced to a small set of physically meaningful parameters, including amplitudes, center times, and pulse widths.
These parameters are optimized with a trust-region method using the PMP-based gradients.

As a concrete example, we study the $|0\rangle\rightarrow |2\rangle$ transfer in a finite-level superconducting transmon driven by two microwave tones.
The higher excited states are treated as explicit leakage states, and energy relaxation is included through an effective non-Hermitian description.
Numerical results show that the optimized pulses preserve the counterintuitive ordering of STIRAP.
They also improve the final transfer fidelity, reduce leakage, shorten the protocol duration, and enlarge the high-fidelity region under amplitude and detuning errors.
These results indicate that PMP-based parameter optimization can provide a practical route to leakage-suppressed STIRAP control in weakly anharmonic multilevel systems.

The remainder of this paper is organized as follows. Sec.~\ref{Sec:Model} introduces the multilevel chain model for STIRAP-type population transfer. Sec.~\ref{Sec:PMP} presents the PMP-based quantum optimal-control framework and derives the gradient formulas for parametrized Gaussian pulses. Sec.~\ref{Sec:numerical} applies the method to a five-level transmon model and reports the numerical results for transfer performance and robustness.

\section{Multilevel chain model for STIRAP-type population transfer}
\label{Sec:Model}
This section introduces the effective multilevel model used throughout this work. We first describe the nearest-neighbor chain Hamiltonian and the associated dark-state structure. We then explain how the same model separates the intended transfer pathway from the leakage manifold in realistic multilevel systems.

\subsection{Effective chain Hamiltonian}

An $N$-state nearest-neighbor chain provides a natural framework for STIRAP-type adiabatic control in multilevel systems. 
In a suitable interaction picture and rotating frame, 
the rotating-wave approximation (RWA) removes rapidly oscillating terms and yields an effective tridiagonal Hamiltonian. 
The diagonal elements represent the effective detunings while the off-diagonal elements describe field-induced couplings between adjacent states in terms of effective Rabi frequencies 
\cite{Kuklinski1989,Gaubatz1990,Bergmann1998,Vitanov2001,Kral2007,Bergmann2015,Vitanov2017}.

In the chain basis $\{|0\rangle,\dots,|N-1\rangle\}$, 
the RWA Hamiltonian is
\begin{equation}
\resizebox{0.15\columnwidth}{!}{$H(t)=\frac{\hbar}{2}$}
\resizebox{0.76\columnwidth}{!}{%
\(
\left[
\begin{array}{cccccc}
2\Delta_0 & \Omega_{0,1}(t) & 0 & \cdots & 0 & 0\\
\Omega_{0,1}^{*}(t) & 2\Delta_1 & \Omega_{1,2}(t) & \cdots & 0 & 0\\
0 & \Omega_{1,2}^{*}(t) & 2\Delta_2 & \cdots & 0 & 0 \\
\vdots & \vdots & \vdots & \ddots & \vdots & \vdots \\
0 & 0 & 0 & \cdots & 2\Delta_{N-2} & \Omega_{N-2,N-1}(t)\\
0 & 0 & 0 & \cdots & \Omega_{N-2,N-1}^{*}(t) & 2\Delta_{N-1}
\end{array}
\right].
\)
}
\label{eq:chain_hamiltonian}
\end{equation}
Here, $\Omega_{j,j+1}(t)$ denotes the generally complex coupling envelope for the transition $|j\rangle \leftrightarrow |j+1\rangle$, 
with both amplitude and phase set by the external drives. 
The quantities $\Delta_j$ are the effective detunings defined by the chosen rotating frame. 
In general, they may be nonzero for both the endpoint and intermediate states.

The relevant multiphoton resonance condition is imposed separately as a constraint on the chosen transfer pathway. 
For an $M$-photon resonance between two selected states, 
the carrier frequencies and the rotating frame are chosen so that the accumulated phase mismatch vanishes. 
Equivalently, the corresponding diagonal mismatch is set to zero in the effective Hamiltonian. 
The usual two-photon resonance condition in three-level STIRAP is the special case $M=2$. 
Under such a resonance condition, the system can support an instantaneous dark eigenstate.

The chain also exhibits an odd-even effect. 
For an odd-length chain, $N=2n+1$, an end-to-end dark state exists when the effective detunings on the populated even sublattice are aligned in the rotating frame.
The corresponding eigenvalue is set by their common detuning and can be shifted to zero by an overall energy offset. 
The amplitudes obey a nearest-neighbor recursion relation and have support only on the even sublattice.
It therefore connects $|0\rangle$ to $|N-1\rangle=|2n\rangle$ under a counterintuitive pulse sequence.

We denote this dark state by $|D_0(t)\rangle$. It can be written as
\begin{equation}
|D_0(t)\rangle
=
\frac{1}{\mathcal N(t)}
\sum_{k=0}^{n} A_k(t)\,|2k\rangle,
\label{eq:chain_dark_state}
\end{equation}
where the normalization factor is
\begin{equation}
\mathcal N(t)
=
\left(
\sum_{k=0}^{n}|A_k(t)|^2
\right)^{1/2},
\label{eq:chain_dark_norm}
\end{equation}
and the coefficients are
\begin{equation}
A_k(t)
=
(-1)^k
\prod_{m=0}^{k-1}\Omega^*_{2m,2m+1}(t)
\prod_{m=k}^{n-1}\Omega_{2m+1,2m+2}(t).
\label{eq:chain_dark_coeff}
\end{equation}
Because $|D_0(t)\rangle$ is supported only on the even sublattice, 
it provides an adiabatic pathway for population transfer from $|0\rangle$ to $|N-1\rangle$. 
By contrast, under the same nearest-neighbor and resonance assumptions, 
an even-length chain does not have the same end-to-end dark-state connectivity. 
Its transport behavior is therefore qualitatively different.

\subsection{Target manifold and leakage manifold}

In realistic multilevel systems, 
the desired target state does not necessarily coincide with the highest state retained in the truncated model. 
The additional states may lie outside the intended transfer pathway,
but they can still affect the dynamics through off-resonant couplings. 
It is therefore useful to separate the Hilbert space into a target manifold, 
which contains the states directly involved in the intended STIRAP-type process, 
and a leakage manifold, 
which contains the remaining higher-lying states.

For a transfer process $|0\rangle \to |m\rangle$ with $m<N-1$, 
we define
\begin{align}
\mathcal H_{\mathrm{tar}}
&=
\mathrm{span}\{|0\rangle,|1\rangle,\dots,|m\rangle\},
\label{eq:Htar}\\
\mathcal H_{\mathrm{leak}}
&=
\mathrm{span}\{|m+1\rangle,\dots,|N-1\rangle\}.
\label{eq:Hleak}
\end{align}
When the target subchain has the appropriate odd-length structure and satisfies the corresponding resonance conditions, the target manifold supports the desired dark-state transport.
The leakage manifold, by contrast, is not part of the ideal STIRAP process. Nevertheless, these higher states remain dynamically relevant because finite anharmonicity and finite-amplitude driving can admix them into the evolution.

Even weak occupation of the leakage manifold can modify the effective transfer process. 
Virtual excitation of the leakage states can shift the effective detunings, 
renormalize the effective couplings, 
and distort the instantaneous dark-state structure. 
A perturbative treatment of these effects is justified when the leakage detunings are much larger than the relevant Rabi frequencies.
When the leakage detunings become comparable to the drive amplitudes, 
the leakage manifold can no longer be eliminated reliably, 
and the multilevel structure must be retained explicitly.

This target--leakage partition is the basis for the optimal-control formulation used below. It allows the desired final-state fidelity and the unwanted leakage population to be included in the same objective functional, so that the pulse optimization can improve state transfer while directly suppressing leakage.

\section{Quantum optimal control framework based on Pontryagin's maximum principle}
\label{Sec:PMP}
This section presents the PMP-based control framework used in this work. 
We first summarize the state-costate structure of PMP, then write its quantum form for Schr\"odinger dynamics, and finally apply it to the multilevel chain to obtain the link-control gradients. 
The detailed PMP derivation is given in Appendix~\ref{app:kkt_pmp}.

\subsection{Control problem and Pontryagin's maximum principle}
We use Pontryagin's maximum principle (PMP) to iteratively update the control pulses in the system Hamiltonian. We first consider a general continuous-time optimal-control problem. The objective functional is written as
\begin{equation}
J[u]=\phi(x(T))+\int_0^T L(x(t),u(t),t)\,\mathrm{d}t,
\end{equation}
where $\phi(x(T))$ and $L(x(t),u(t),t)$ are the terminal cost and the running cost, respectively. Here, $x(t)\in\mathbb{R}^n$ is the system state, and $u(t)\in\mathbb{R}^m$ is the control.

The system dynamics are described by
\begin{equation}
\dot{x}(t)=f(x(t),u(t),t),\quad t\in[0,T].
\end{equation}
In PMP, the dynamical equation is treated as an infinite-dimensional equality constraint. We introduce a costate $\lambda(t)\in\mathbb{R}^n$. The costate satisfies
\begin{equation}
\begin{aligned}
\dot{\lambda}^{T}(t)
% &=
% -\frac{\partial \mathcal{H}_{\mathrm{P}}}{\partial x}(x,u,\lambda,t)\\
=
\frac{\partial L}{\partial x}(x,u,t)-\lambda^{T}(t)\frac{\partial f}{\partial x}(x,u,t),
\end{aligned}
\end{equation}
with the boundary conditions
\begin{equation}
x(0)=x_0,\quad \lambda^{T}(T)=-\frac{\partial \phi}{\partial x}(x(T)).
\end{equation}
The Pontryagin Hamiltonian is defined as
\begin{equation}
\mathcal{H}_{\mathrm{P}}(x,u,\lambda,t)=\lambda^{T} f(x,u,t)-L(x,u,t).
\end{equation}
The evolution equation for $\lambda(t)$ is also called the adjoint equation. It propagates the influence of the terminal objective and the running cost backward in time. Therefore, in numerical implementations, $\lambda(t)$ is integrated from $t=T$ to $t=0$.

The Pontryagin Hamiltonian gives the continuous-time functional gradient of the optimal-control problem with respect to the control pulse:
\begin{equation}
\begin{aligned}
\frac{\delta J}{\delta u(t)}
&=
-\frac{\partial \mathcal{H}_{\mathrm P}}{\partial u}(x(t),u(t),\lambda(t),t)\\
&=
\frac{\partial L}{\partial u}(x(t),u(t),t)
-
\lambda^{T}(t)\frac{\partial f}{\partial u}(x(t),u(t),t).
\end{aligned}
\end{equation}
This control gradient is the central quantity in the iterative update. Further details of the PMP formalism are given in  Appendix~\ref{app:kkt_pmp}.

\subsection{PMP optimization for quantum dynamics}
\label{Sec:Opt_Quantum}

In quantum control, the system state is a quantum state in a complex Hilbert space. We use a complex formulation that is equivalent to the real-variable PMP. For a closed quantum system governed by Schr\"odinger dynamics,
\begin{equation}
i\hbar\frac{\mathrm{d}}{\mathrm{d}t}|\psi(t)\rangle = H(t;u)|\psi(t)\rangle,
\end{equation}
we assume that the Hamiltonian is linear in the controls:
\begin{equation}
H(t;u)=H_0+\sum_i u_i(t)H_i,
\end{equation}
where $u_i(t)\in\mathbb{R}$ are real control fields. The real-valued Pontryagin Hamiltonian can then be defined as
\begin{equation}
\resizebox{0.88\columnwidth}{!}{$\displaystyle
\mathcal{H}_{\mathrm{P}}(\psi,\lambda,u,t)
=
\mathrm{Im}\!\left\{\frac{1}{\hbar}\langle\lambda(t)|H(t;u)|\psi(t)\rangle\right\}
-
L(\psi,u,t).
$}
\end{equation}
Here, $\langle\lambda(t)|$ denotes the costate. The corresponding adjoint equation and terminal condition are
\begin{align}
&\frac{\mathrm{d}}{\mathrm{d}t}\langle\lambda(t)|
=
\frac{i}{\hbar}\langle\lambda(t)|H(t;u)
+
\frac{\partial L}{\partial \ket{\psi(t)}},\\
&\langle\lambda(T)|
=
-\frac{\partial \phi}{\partial \ket{\psi(T)}}.
\end{align}
The continuous-time gradient with respect to each control field is
\begin{equation}
\frac{\delta J}{\delta u_i(t)}
=
\frac{\partial L}{\partial u_i}(\psi(t),u(t),t)
-
\frac{1}{\hbar}\mathrm{Im}\!\left\{
\langle\lambda(t)| H_i |\psi(t)\rangle
\right\}.
\end{equation}
After time discretization, 
the resulting adjoint gradient has the same forward--backward structure as the standard GRAPE method, 
which combines forward propagation of the quantum state with backward propagation of an adjoint variable \cite{Khaneja2005GRAPE,PRXQuantum.2.030203}.
Therefore, the two approaches share the same basic forward--backward structure.

\subsection{PMP gradients for the multilevel chain}
For the $N$-state nearest-neighbor chain Hamiltonian introduced in the previous section, the control problem can be written in a form naturally suited to PMP. We decompose each complex nearest-neighbor coupling as
$\Omega_{j,j+1}(t)=u_j^{(x)}(t)+i\,u_j^{(y)}(t)$.
The Hamiltonian then becomes
\begin{equation}
H(t)=H_{\mathrm d}+\sum_{j=0}^{N-2}\Bigl(u_j^{(x)}(t)X_j+u_j^{(y)}(t)Y_j\Bigr),
\end{equation}
where
\begin{align}
&H_{\mathrm d}=\hbar\sum_{j=0}^{N-1}\Delta_j|j\rangle\langle j|,\\
&X_j=\frac{\hbar}{2}\Bigl(|j\rangle\langle j+1|+|j+1\rangle\langle j|\Bigr),\\
&Y_j=\frac{\hbar}{2}\Bigl(i|j\rangle\langle j+1|-i|j+1\rangle\langle j|\Bigr).
\end{align}
Thus, the optimal-control problem is reduced to the joint optimization of the link controls $u_j^{(x)}(t)$ and $u_j^{(y)}(t)$ under the dynamical constraint.

In this representation, PMP gives the following control gradients:
\begin{align}
&\frac{\delta J}{\delta u_j^{(x)}(t)}
=
\frac{\partial L}{\partial u_j^{(x)}}
-
\frac{1}{\hbar}
\operatorname{Im}\!\left\{
\langle\lambda(t)| X_j |\psi(t)\rangle
\right\},\\
&\frac{\delta J}{\delta u_j^{(y)}(t)}
=
\frac{\partial L}{\partial u_j^{(y)}}
-
\frac{1}{\hbar}
\operatorname{Im}\!\left\{
\langle\lambda(t)| Y_j |\psi(t)\rangle
\right\}.
\end{align}
The control update therefore depends on both the forward-propagated quantum state and the backward-propagated costate. 
This reflects the optimization structure of the nearest-neighbor chain model.
The functional gradients derived here are independent of the specific pulse ansatz. 
In Sec.~\ref{Sec:numerical}, they are projected onto a finite set of Gaussian pulse parameters to keep the optimization experimentally interpretable and STIRAP-like.

This formulation is especially suitable for STIRAP-type adiabatic transfer. 
In the chain model, dark-state preservation, nonadiabatic deviation, 
and leakage suppression can all be represented by subspace projections. 
They can then be incorporated into a unified objective functional. 
For the multistate-chain STIRAP problem considered here, 
the same PMP framework can also be extended to effective non-Hermitian dynamics and to more general open-system master-equation dynamics.
It is therefore well suited to multi-objective control problems involving both fidelity optimization and leakage suppression.

\section{Application to optimal STIRAP in a transmon}
\label{Sec:numerical}
This section applies the PMP-based framework to a weakly anharmonic superconducting transmon. 
The aim is to optimize a STIRAP-type transfer from $|0\rangle$ to $|2\rangle$ while explicitly suppressing leakage to nearby higher levels. 
We first derive the  effective five-level Hamiltonian and identify the leakage mechanism. 
We then introduce the constrained Gaussian pulse family and the leakage-penalized objective functional. 
Finally, we compare the initial and PMP-optimized protocols and analyze their robustness against parameter imperfections.

\subsection{Five-level transmon model and leakage mechanism}
\label{Sec:transmon_model}

A weakly anharmonic transmon provides a concrete setting where the ideal three-level STIRAP picture is only approximate. 
The target process is the population transfer $|0\rangle \rightarrow |2\rangle$. 
However, finite-amplitude driving can also couple the target manifold to nearby higher excited states. 
These off-target transitions are the dominant source of leakage in the present model.

\begin{figure}
    \centering
    \includegraphics[width=1\linewidth]{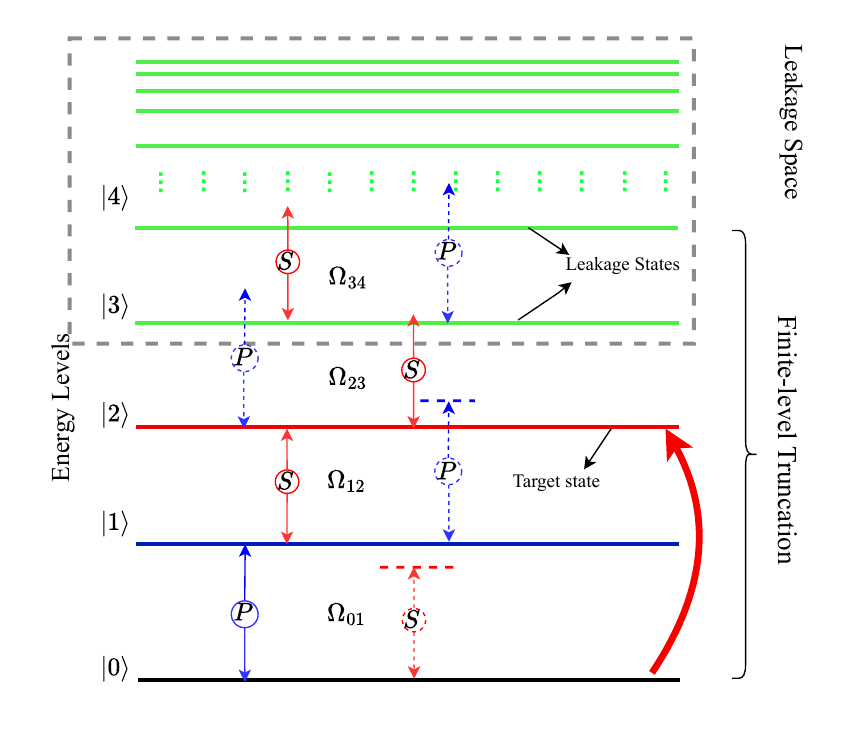}
    \caption{Five-level truncation of a transmon ladder for leakage-suppressed STIRAP control. 
    The target process is the population transfer $|0\rangle \rightarrow |2\rangle$, whereas $|3\rangle$ and $|4\rangle$ are treated as explicit leakage states. 
    The adjacent couplings $\Omega_{01}$, $\Omega_{12}$, $\Omega_{23}$, and $\Omega_{34}$ define the effective nearest-neighbor chain. 
    Higher levels beyond the five-level truncation are not included explicitly.
    The $P$ and $S$ labels identify the pump and Stokes tones, respectively. Solid arrows indicate the intended near-resonant couplings, whereas dashed arrows denote off-resonant couplings induced by the same tones.}
    \label{fig:xmon}
\end{figure}

We truncate the transmon Hilbert space to the five-level basis 
$\{|0\rangle,\dots,|4\rangle\}$. 
The target manifold is
$\mathcal H_{\mathrm{tar}}=\mathrm{span}\{|0\rangle,|1\rangle,|2\rangle\}$, 
whereas the leakage manifold is
$\mathcal H_{\mathrm{leak}}=\mathrm{span}\{|3\rangle,|4\rangle\}$. 
The effective chain structure is shown in Fig.~\ref{fig:xmon}. 
This finite-level model captures both the intended STIRAP pathway and the leading leakage channels within the RWA description 
\cite{Kumar2016,Xu2016,Vepsalainen2019,Zheng2022npjQI_OCT_STIRAP_Qudit,Niu2022PRAppSTIRUP}.
Further details on the transmon are given in Appendix~\ref{app:transmon}.

The drift Hamiltonian is
\begin{equation}
H_0=\sum_{n=0}^{4}E_n|n\rangle\langle n|,
\label{eq:H0}
\end{equation}
where $E_n$ are the eigenenergies of the transmon circuit. 
The control field is introduced through capacitive coupling to the Cooper-pair number operator $\hat n$,
\begin{equation}
H_c(t)=2e\,V(t)\,\hat n.
\label{eq:Hd}
\end{equation}
To implement the two-pulse structure of STIRAP, we use a two-tone drive,
\begin{equation}
V(t)=V_p(t)\cos(\omega_p t+\phi_p)+V_s(t)\cos(\omega_s t+\phi_s),
\label{eq:V_two_tone}
\end{equation}
where $p$ and $s$ denote the pump and Stokes components. 
The functions $V_p(t)$ and $V_s(t)$ are the slowly varying pulse envelopes.

After applying a suitable unitary transformation and RWA, the driven transmon is reduced to the effective nearest-neighbor Hamiltonian
\begin{equation}
\tilde{H}(t)=
\frac{\hbar}{2}
\begin{pmatrix}
0 & \Omega_{01}(t) & 0 & 0 & 0\\
\Omega_{01}^*(t) & 2\Delta_1 & \Omega_{12}(t) & 0 & 0\\
0 & \Omega_{12}^*(t) & 2\Delta_2 & \Omega_{23}(t) & 0\\
0 & 0 & \Omega_{23}^*(t) & 2\Delta_3 & \Omega_{34}(t)\\
0 & 0 & 0 & \Omega_{34}^*(t) & 2\Delta_4
\end{pmatrix}.
\label{eq:H_chain_main}
\end{equation}
The effective detunings are
\begin{equation}
\Delta_1=\omega_{10}-\omega_p,
\label{eq:Delta1}
\end{equation}
\begin{equation}
\Delta_2=\omega_{20}-(\omega_p+\omega_s)\equiv \delta,
\label{eq:Delta2}
\end{equation}
\begin{equation}
\Delta_3=\omega_{30}-(2\omega_p+\omega_s),
\label{eq:Delta3}
\end{equation}
\begin{equation}
\Delta_4=\omega_{40}-(2\omega_p+2\omega_s).
\label{eq:Delta4}
\end{equation}
Here $\omega_{jk}=(E_j-E_k)/\hbar$, and $\delta$ is the two-photon detuning of the target STIRAP process. 
The nominal working point satisfies
\begin{equation}
\delta=\omega_{20}-(\omega_p+\omega_s)=0,
\label{eq:two_photon_resonance}
\end{equation}
while small one-photon detunings are allowed and are subsequently scanned to assess robustness.

We include energy relaxation through an effective non-Hermitian Hamiltonian,
\begin{equation}
H_{\mathrm{nh}}(t)=\tilde{H}(t)-\frac{i\hbar}{2}\sum_{\mu}C_\mu^\dagger C_\mu,
\label{eq:Hnh}
\end{equation}
where the collapse operators $C_\mu$ represent the dominant relaxation channels. 
The state then evolves according to
\begin{equation}
i\hbar\frac{d}{dt}|\psi(t)\rangle = H_{\mathrm{nh}}(t)|\psi(t)\rangle.
\label{eq:state_nh}
\end{equation}
This effective no-jump description incorporates relaxation-induced loss into the state propagation used in the optimization.

The control-induced couplings inherit the nearest-neighbor matrix elements of the weakly anharmonic transmon. 
Within the weakly anharmonic oscillator approximation, they are written as
\begin{equation}
\begin{aligned}
&\Omega_{01}(t)=\Omega_p(t)e^{i\phi_p},\quad
\Omega_{12}(t)=\sqrt{2}\,\Omega_s(t)e^{i\phi_s},\\
&\Omega_{23}(t)=\sqrt{3}\,\Omega_p(t)e^{i\phi_p},\quad
\Omega_{34}(t)=\sqrt{4}\,\Omega_s(t)e^{i\phi_s}.
\end{aligned}
\end{equation}
Here, $\Omega_{p/s}(t) \propto 2e n_{\mathrm{zpf}} V_{p/s}(t)$ are the effective Rabi envelopes, where $n_{\mathrm{zpf}}$ denotes the zero-point-fluctuation amplitude of the Cooper-pair number operator.
The same two tones that drive the target transitions $|0\rangle \leftrightarrow |1\rangle$ and $|1\rangle \leftrightarrow |2\rangle$ also drive the higher transitions $|2\rangle \leftrightarrow |3\rangle$ and $|3\rangle \leftrightarrow |4\rangle$. 
This is the central mechanism by which leakage enters the multilevel STIRAP problem.

If the higher couplings $\Omega_{23}$ and $\Omega_{34}$ are neglected, the target three-level manifold has the usual instantaneous dark state
\begin{equation}
|D_{\mathrm{tar}}(t)\rangle=
\frac{\Omega_{12}(t)|0\rangle-\Omega_{01}(t)|2\rangle}
{\sqrt{|\Omega_{01}(t)|^2+|\Omega_{12}(t)|^2}}.
\label{eq:D_tar}
\end{equation}
With the mixing angle $\theta(t)$
\begin{equation}
\tan\theta(t)=\frac{|\Omega_{01}(t)|}{|\Omega_{12}(t)|},
\label{eq:mixing_angle}
\end{equation}
this state can also be written as
\begin{equation}
|D_{\mathrm{tar}}(t)\rangle=
\cos\theta(t)\,|0\rangle-e^{i\varphi(t)}\sin\theta(t)\,|2\rangle,
\label{eq:D_tar_theta}
\end{equation}
where $\varphi(t)=-\arg\Omega_{01}(t)-\arg\Omega_{12}(t)$. 
In the standard counterintuitive sequence, the Stokes pulse precedes the pump pulse. 
Then $\theta(t)$ changes smoothly from $0$ to $\pi/2$, and adiabatic following of $|D_{\mathrm{tar}}(t)\rangle$ transfers the population from $|0\rangle$ to $|2\rangle$ 
\cite{Bergmann1998,Vitanov2001,Kral2007,Bergmann2015}. 
In the full five-level model, however, this target dark state is no longer an exact eigenstate. 
Higher-level couplings introduce leakage and deform the dark-state pathway.

\subsection{Gaussian pulse parametrization and optimization objective}
\label{Sec:gaussian_objective}
We now restrict the controls to a Gaussian STIRAP pulse family. 
This choice keeps the protocol smooth, bandwidth-limited, and experimentally transparent. 
It also reduces the optimization to a small number of pulse parameters with direct physical meaning.

This low-dimensional parametrization is also well suited to the trust-region optimization employed below, which uses a Broyden–Fletcher–Goldfarb–Shanno (BFGS) approximation, because it keeps the curvature approximation tractable and avoids the large overhead associated with time-sliced control variables. Further details on the computational cost of this parametrized optimization are given in Appendix~\ref{app:trust_region_bfgs} and Appendix~\ref{App:complexity}.

The pump and Stokes envelopes are
\begin{equation}
\Omega_{p/s}(t)=A_{p/s}\exp\left[-\frac{(t-t_{0p/s})^2}{2\sigma_{p/s}^2}\right].
\label{eq:gaussian_pulses}
\end{equation}
The optimization variables are
\begin{equation}
\mathbf{u}=
(A_p,A_s,t_{0p},t_{0s},\sigma_p,\sigma_s).
\label{eq:alpha_vec}
\end{equation}
Here $A_p$ and $A_s$ set the peak amplitudes, $t_{0p}$ and $t_{0s}$ set the pulse centers, and $\sigma_p$ and $\sigma_s$ set the pulse widths. 
These six parameters control the pulse overlap, relative delay, peak strength, and bandwidth. 
Thus, the ansatz is restrictive but physically interpretable. 
Accordingly, the optimized pulse should be understood as a local solution obtained through PMP-based optimization within the constrained Gaussian STIRAP family, rather than as a globally optimal waveform over the unrestricted control space.
It is designed to optimize a STIRAP-like protocol, not to search over arbitrary waveforms.

For an initial state $|0\rangle$, we define the Bolza-type objective functional
\begin{equation}
\begin{aligned}
J(\mathbf{u})
=
&w_f(1-|\langle 2|\psi(T;\mathbf{u})\rangle|^2) \\
&+
\int_0^T
\left[
w_1 P_1(t)
+
w_{\mathrm{leak}}\bigl(P_3(t)+P_4(t)\bigr)
\right]dt.
\label{eq:bolza_transmon}
\end{aligned}
\end{equation}
Here
$P_n(t)=|\langle n|\psi(t)\rangle|^2$
is the instantaneous weight of level $|n\rangle$. 
Because $H_{\mathrm{nh}}$ is non-Hermitian, these quantities are unnormalized weights in the effective no-jump evolution. 
The terminal term penalizes target-state infidelity. 
The running cost penalizes population in the intermediate state $|1\rangle$ and in the leakage states $|3\rangle$ and $|4\rangle$. 
This choice reflects the main error channels of multilevel STIRAP in a weakly anharmonic device 
\cite{Yatsenko2014,PRXQuantum.4.030312}. 
Population in $|1\rangle$ indicates a departure from the dark-state pathway, while population in $|3\rangle$ and $|4\rangle$ measures leakage caused by off-resonant excitation.

The PMP gradients derived in Sec.~\ref{Sec:PMP} are functional gradients with respect to the control fields. 
For the Gaussian ansatz, they are converted into parameter gradients through
\begin{equation}
\frac{\partial J}{\partial u_k}
=
\int_0^T
\left[
\frac{\delta J}{\delta \Omega_p(t)}
\frac{\partial \Omega_p(t)}{\partial u_k}
+
\frac{\delta J}{\delta \Omega_s(t)}
\frac{\partial \Omega_s(t)}{\partial u_k}
\right]dt,
\label{eq:param_grad}
\end{equation}
where $u_k\in\{A_p,A_s,t_{0p},t_{0s},\sigma_p,\sigma_s\}$. The required Gaussian derivatives are analytic.
\begin{align}
&\frac{\partial \Omega_\mu(t)}{\partial A_\mu}
=
\frac{\Omega_\mu(t)}{A_\mu},
\qquad
\mu\in\{p,s\},\\
&\frac{\partial \Omega_\mu(t)}{\partial t_{0\mu}}
=
\Omega_\mu(t)\frac{t-t_{0\mu}}{\sigma_\mu^2},
\qquad
\mu\in\{p,s\},\\
&\frac{\partial \Omega_\mu(t)}{\partial \sigma_\mu}
=
\Omega_\mu(t)\frac{(t-t_{0\mu})^2}{\sigma_\mu^3},
\qquad
\mu\in\{p,s\}.
\end{align}
The PMP iteration therefore yields a direct gradient-based optimization over a small set of pulse parameters with clear experimental interpretation \cite{Liberzon2012COVandOCT,PRXQuantum.2.030203}.
Thus, the present parametrization does not require independent optimization of the pulse amplitude at every time step, although the PMP framework itself could be extended to such time-sliced controls.

We initialize the optimization with a conventional counterintuitive Gaussian STIRAP pair, in which the Stokes pulse precedes the pump pulse. 
This initial guess already produces reasonable transfer within the target manifold. 
Its fidelity is nevertheless limited by the leakage channels described above. 
The PMP update then adjusts the amplitudes, centers, and widths to reduce the leakage penalty and improve the final transfer. 
The numerical simulations were implemented using the Quantum Toolbox in Python (QuTiP) and related Python-based tools
\cite{Johansson2012,Lambert2026PhysRepQuTiP5}.
This initialization is chosen to preserve the dark-state boundary conditions of STIRAP, and the resulting protocol is therefore a constrained local optimum connected to the STIRAP transfer mechanism.

\subsection{Transfer performance and robustness}
\label{Sec:performance_robustness}
We now compare the initial Gaussian STIRAP protocol with the PMP-optimized protocol. 
The main result is that the optimized pulse pair keeps the counterintuitive STIRAP ordering, but achieves higher fidelity, lower leakage, and a broader high-fidelity region under parameter variations.

\begin{figure}
    \centering
    \includegraphics[width=1\linewidth]{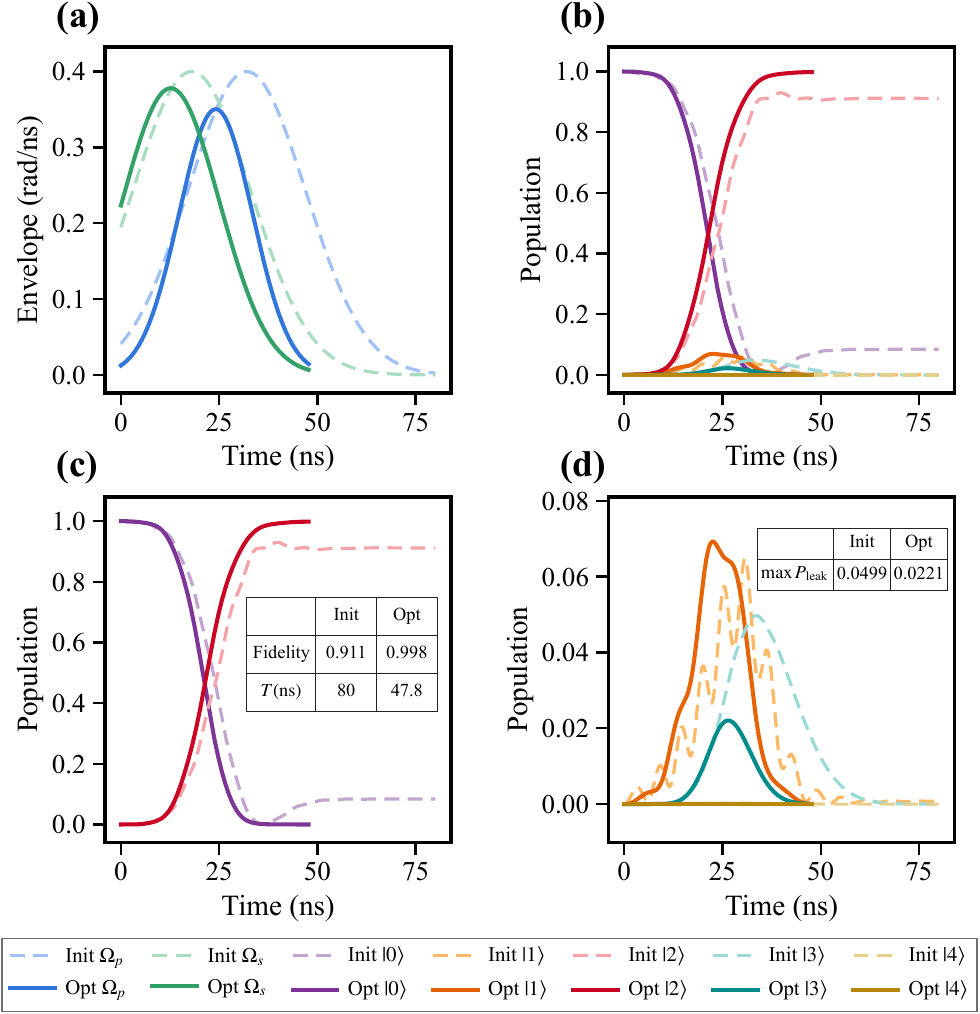}
    \caption{Initial and PMP-optimized Gaussian protocols for the five-level transmon chain. (a) Pump and Stokes envelopes for the initial (dashed) and optimized (solid) pulses. The optimized protocol retains the counterintuitive STIRAP ordering while substantially shortening the total duration. (b),(c) Time-dependent level populations for the initial and optimized protocols, respectively. Relative to the initial protocol, the optimized pulses increase the final population of the target state \(|2\rangle\) and reduce transient occupation of the penalty states \(|1\rangle\), \(|3\rangle\), and \(|4\rangle\). (d) Leakage population for the two protocols. Insets report the transfer fidelity, total pulse duration, and maximum leakage population.}
    \label{fig:pulse_state}
\end{figure}

Figure~\ref{fig:pulse_state} shows the pulse envelopes and the corresponding population dynamics. 
The optimized protocol still places the Stokes pulse before the pump pulse.
Thus, the optimized protocol remains STIRAP-like in its pulse ordering and overall transfer pathway.
Instead, it reshapes the pulse widths, overlap, and relative timing within the Gaussian family. 
As shown in Fig.~\ref{fig:pulse_state}(a), the total duration is reduced from $80~\mathrm{ns}$ to $47.8~\mathrm{ns}$.

\begin{figure*}
    \centering
    \includegraphics[width=1\linewidth]{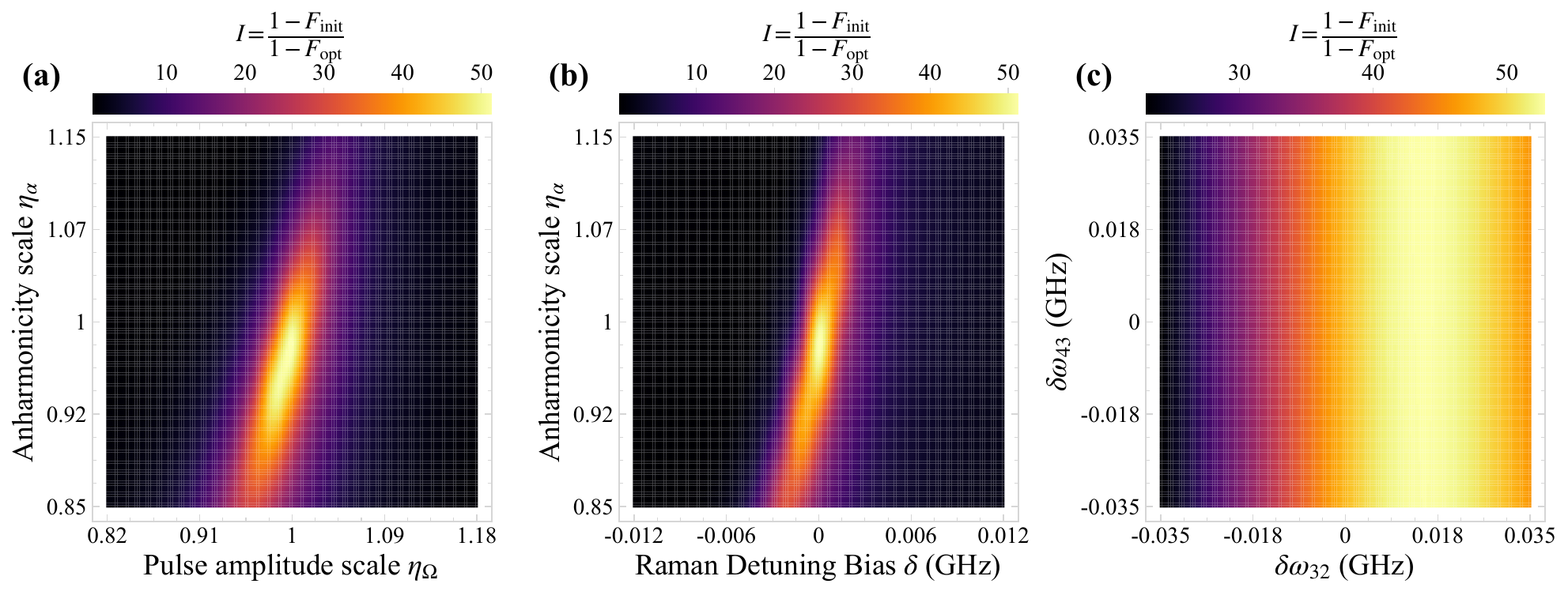}
    \caption{Improvement landscapes for the PMP-optimized protocol relative to the initial Gaussian protocol, quantified by $I=(1-F_{\mathrm{init}})/(1-F_{\mathrm{opt}})$, over representative parameter planes. (a) Pulse-amplitude scale $\eta_\Omega$ versus anharmonicity scale $\eta_\alpha$. (b) Raman-detuning bias $\delta$ versus anharmonicity scale $\eta_\alpha$. (c) Higher-transition frequency offsets $\delta\omega_{32}$ and $\delta\omega_{43}$. Brighter colors correspond to larger improvement factors and hence stronger suppression of transfer error by the optimized protocol.}
    \label{fig:improvement_scan}
\end{figure*}

The transfer performance is also improved substantially. 
The final target-state fidelity increases from $0.911$ to $0.998$. 
At the same time, the maximum leakage population,
\begin{equation}
P_{\mathrm{leak}}(t)=P_3(t)+P_4(t),
\label{eq:Perror}
\end{equation}
is reduced from $0.0499$ to $0.0221$. 
The population traces show that the optimized pulse also reduces transient occupation of $|1\rangle$, $|3\rangle$, and $|4\rangle$. 
The optimized evolution therefore stays closer to the desired dark-state transfer path while avoiding the leakage manifold.

The simultaneous increase in fidelity and reduction in pulse duration are not simply explained by a trivial speedup.
A shorter pulse usually has a broader spectrum and can enhance unwanted transitions in a multilevel ladder. 
The improvement therefore comes from a better balance between finite-time adiabatic following and leakage suppression. 
The PMP update finds this balance by adjusting the pulse overlap and relative timing within the constrained Gaussian family. 
The adiabaticity analysis in Appendix~\ref{App:adiabaticity} shows that the optimized protocol is better interpreted as leakage-suppressed finite-time dressed-state transfer rather than as a simple improvement of ideal three-level adiabaticity.
This picture is consistent with previous studies of dissipative chain transfer, which show that the optimality of STIRAP depends on the topology of the chain and on finite-power constraints \cite{PhysRevA.85.033417}. 

To quantify robustness, we compare the optimized protocol and the initial protocol under parameter perturbations. 
We define the improvement factor
\begin{equation}
I=\frac{1-F_{\mathrm{init}}}{1-F_{\mathrm{opt}}},
\label{eq:improvement_factor}
\end{equation}
where $F_{\mathrm{init}}$ and $F_{\mathrm{opt}}$ are the final target-state fidelities under the same perturbed condition. 
Values $I>1$ mean that the optimized protocol reduces the transfer error. 
Larger values indicate stronger improvement.
\begin{figure}
    \centering
    \includegraphics[width=1\linewidth]{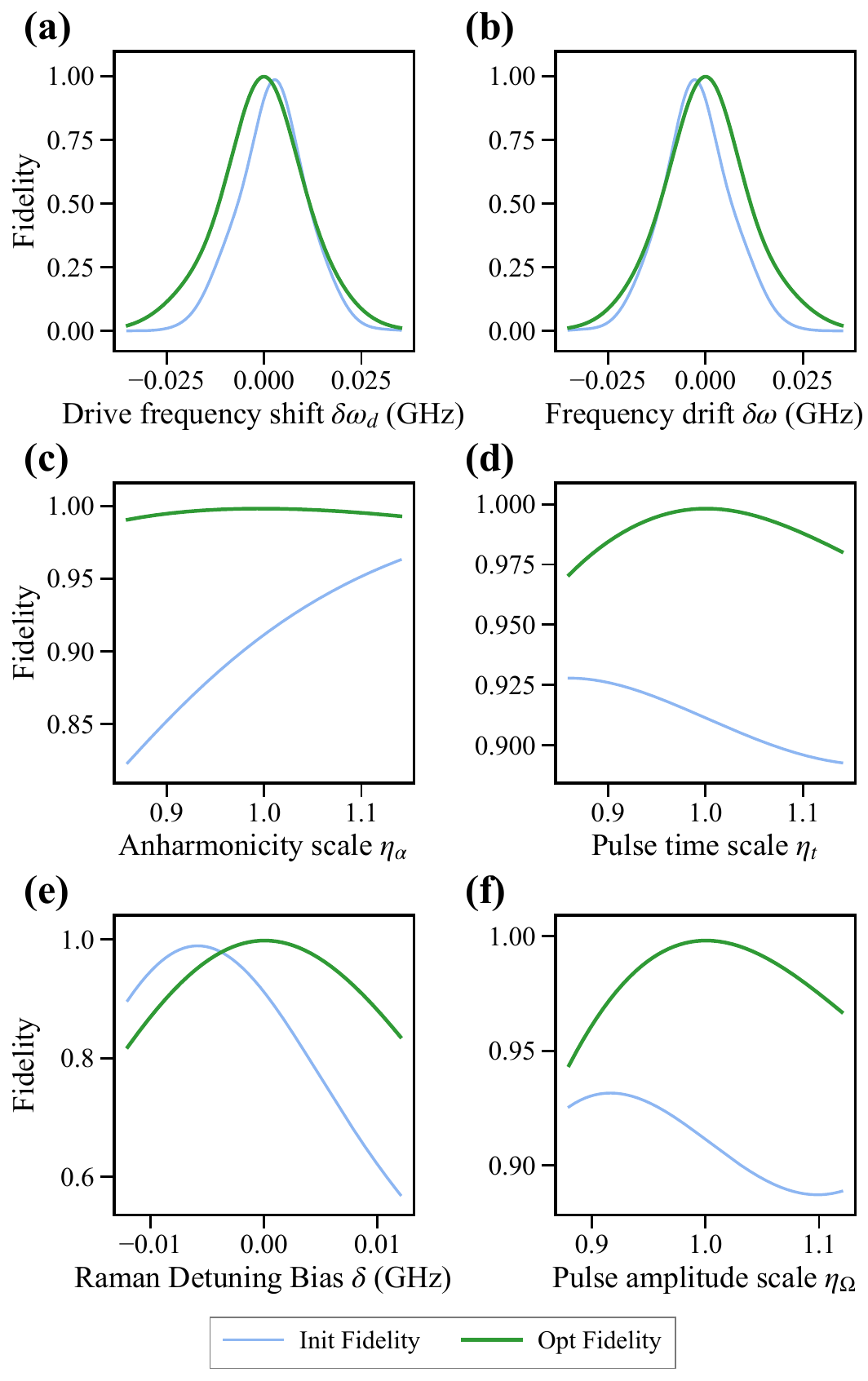}
    \caption{One-dimensional robustness cuts for the initial and PMP-optimized protocols. The final target-state fidelity is plotted versus (a) the global drive-frequency shift $\delta\omega_d$, (b) the common transition-frequency drift $\delta\omega$, (c) the anharmonicity scale $\eta_\alpha$, (d) the pulse-time scale $\eta_t$, (e) the Raman-detuning bias $\delta$, and (f) the pulse-amplitude scale $\eta_\Omega$. Relative to the initial protocol, the optimized protocol retains a broader high-fidelity region across all scans.}
    \label{fig:robustness}
\end{figure}

Figure~\ref{fig:improvement_scan} shows two-dimensional improvement landscapes. 
In Fig.~\ref{fig:improvement_scan}(a), the pulse-amplitude scale $\eta_\Omega$ and the anharmonicity scale $\eta_\alpha$ are varied together. 
A broad high-improvement region appears around the nominal working point. 
This means that the optimized protocol remains advantageous when the drive strength and level anharmonicity change simultaneously. 
These two parameters jointly determine how strongly the higher transitions are activated and how much the target dark-state pathway is distorted.

Figure~\ref{fig:improvement_scan}(b) scans the Raman detuning bias $\delta$ and the anharmonicity scale $\eta_\alpha$. 
The high-improvement region forms a tilted band near the nominal working point. 
This behavior shows that the optimized protocol is less sensitive to combined errors in the two-photon resonance condition and the anharmonicity. 
Both errors deform the dark-state pathway and increase off-resonant admixture. 
The simultaneous suppression of their effects indicates that the optimization targets the relevant multilevel leakage mechanism.

Figure~\ref{fig:improvement_scan}(c) probes uncertainty in the higher transition frequencies through $\delta\omega_{32}$ and $\delta\omega_{43}$. 
The improvement factor remains high over a broad region. 
It depends more weakly on $\delta\omega_{43}$ than on $\delta\omega_{32}$, which corresponds to the first leakage transition adjacent to the target manifold.
This is physically reasonable. 
In a weakly anharmonic ladder, the nearest leakage level usually gives the strongest distortion of the target STIRAP pathway.
This behavior also suggests that the convergence with respect to the truncation size is governed mainly by the detuning of the nearest leakage levels, while more distant levels give smaller corrections when they remain spectrally well separated.

We also perform one-dimensional robustness cuts around the nominal point. 
The varied parameters are the global drive-frequency shift $\delta\omega_d$, the common transition-frequency drift $\delta\omega$, the anharmonicity scale $\eta_\alpha$, the pulse-time scale $\eta_t$, the Raman detuning bias $\delta$, and the pulse-amplitude scale $\eta_\Omega$. 
For the time scaling, we rescale the pulse centers and widths as
\begin{equation}
t_{0\mu}\rightarrow \eta_t\,t_{0\mu},
\qquad
\sigma_\mu\rightarrow \eta_t\,\sigma_\mu,
\qquad
\mu\in\{p,s\}.
\label{eq:time_scaling}
\end{equation}
The amplitude scaling is
\begin{equation}
\Omega_\mu(t)\rightarrow \eta_\Omega\,\Omega_\mu(t),
\qquad
\mu\in\{p,s\}.
\label{eq:amp_scaling}
\end{equation}

The fidelity curves are shown in Fig.~\ref{fig:robustness}. 
Figures~\ref{fig:robustness}(a) and \ref{fig:robustness}(b) show that 
the optimized protocol retains a broader high-fidelity region under drive-frequency offsets and transition-frequency drifts. 
It is therefore less sensitive to microwave calibration errors and slow device-frequency fluctuations. 
This enlarged tolerance window is useful for repeated experimental calibration.

Figures~\ref{fig:robustness}(c) and \ref{fig:robustness}(d) show the response to anharmonicity and pulse-time variations. 
In both scans, the optimized protocol outperforms the initial one over the whole range. 
In particular, the optimized fidelity remains close to unity over a wide interval of $\eta_\alpha$. 
The initial protocol deteriorates more rapidly when the device deviates from the nominal anharmonicity. 
This supports the physical picture that the optimized pulses are not merely tuned to one ideal parameter point. 
Instead, they reshape the transfer pathway so that it is less vulnerable to multilevel distortions. 
The improved tolerance to pulse-time rescaling also shows that the optimized solution can withstand moderate global timing-scale errors.

The strongest contrast appears in the Raman-detuning scan in Fig.~\ref{fig:robustness}(e). 
The initial protocol shows a clear asymmetry and loses fidelity rapidly when the Raman bias moves away from resonance. 
The optimized protocol keeps a broader and more symmetric high-fidelity region. 
This indicates that the leakage-suppressed PMP optimization compensates, at least partially, for the detuning-induced deformation of the target STIRAP process. 
A similar improvement appears in the amplitude scan in Fig.~\ref{fig:robustness}(f). 
The optimized pulse is less sensitive to amplitude miscalibration.

To compare the control resources required by the two protocols, we also evaluate the pulse fluence

\begin{equation}
C_{\Omega}
=
\int_0^T
\left[
|\Omega_p(t)|^2+|\Omega_s(t)|^2
\right]dt .
\label{eq:pulse_fluence}
\end{equation}
Since the Rabi envelopes are proportional to the applied microwave voltage envelopes, 
\(C_{\Omega}\) serves as a calibration-dependent proxy for the time-integrated classical drive power.
For the initial protocol, we obtain \(C_{\Omega}^{\mathrm{init}}=8.312\), whereas the PMP-optimized protocol gives \(C_{\Omega}^{\mathrm{opt}}=4.937\).
At the same time, the final fidelity increases from \(0.911\) to \(0.998\), and the maximum leakage population decreases from \(0.0499\) to \(0.0221\).
Thus, the optimized pulse improves fidelity and suppresses leakage without increasing the total pulse fluence.
Here \(C_{\Omega}\) provides a practical control-cost measure for comparing fidelity, duration, and leakage.
In this limited sense, the present fidelity--time--fluence comparison is connected to broader cost--performance trade-offs discussed in quantum information engines and quantum-battery charging~\cite{Kirchberg2025QuantumInformationEngines,Hagman2025ParetoInformationEngines,Hovhannisyan2020ChargingThermalization}.

Overall, the numerical results show that the PMP-based framework provides a practical route to leakage-suppressed STIRAP control in weakly anharmonic superconducting circuits. 
The optimized Gaussian pulses preserve the counterintuitive STIRAP structure. 
They also improve the final transfer fidelity, shorten the protocol, reduce transient occupation of the intermediate and leakage states, and enlarge the high-fidelity operating region under parameter uncertainty. 
More broadly, these results show that embedding STIRAP in a constrained multilevel optimal-control framework can retain the robustness of adiabatic passage while mitigating leakage in realistic multilevel devices.

\section{Conclusions and outlook}
\label{Sec:conclusion}
In this work, we developed a PMP-based optimal-control framework for STIRAP-type population transfer in multilevel quantum systems. By embedding the target process in an \(N\)-state nearest-neighbor chain, the method incorporates both the intended transfer pathway and the dominant leakage channels within a unified effective description. It thereby captures how off-target couplings, finite anharmonicity, and dissipation degrade dark-state transfer in realistic multilevel settings. Within this formulation, pulse design is cast as a leakage-penalized Bolza-type optimal-control problem. The corresponding state-costate equations and control gradients were derived in a complex Hilbert-space representation fully equivalent to the standard real-variable PMP. Restricting the controls to parametrized Gaussian pump and Stokes pulses reduces the original functional optimization problem to a low-dimensional search over experimentally meaningful pulse parameters, while still allowing fidelity, leakage suppression, and robustness to be optimized within a single framework.

As a concrete application, we studied the \(|0\rangle \rightarrow |2\rangle\) transfer problem in a five-level superconducting transmon model with relaxation incorporated through an effective non-Hermitian description. The optimized pulses preserve the counterintuitive ordering characteristic of STIRAP. They also improve the transfer fidelity, suppress transient occupation of intermediate and leakage states, and enlarge the high-fidelity operating region under amplitude miscalibration and detuning errors. These results establish the present framework as a practical route to leakage-suppressed STIRAP control with enhanced robustness in weakly anharmonic multilevel superconducting circuits. Our results also show that leakage is mainly governed by the nearest leakage transitions. Thus, the relevant scaling is set by leakage detunings relative to the drive bandwidth, rather than by the truncation size alone. Related discrete-to-continuous crossover effects have been discussed in biased random-walk models~\cite{Kirchberg2023Entropy}.

From a broader optimal-control perspective, the present PMP-based framework complements Krotov-type methods that offer monotonic convergence~\cite{Reich2012KrotovJCP}. It also connects earlier analyses of adiabatic population transfer and optimal control in dissipative chains~\cite{PhysRevA.85.033417} with optimal-control studies of superconducting circuit-QED platforms~\cite{Goerz2017Charting}. Several natural extensions remain. Full open-system formulations could be developed using Lindblad master equations and dissipative optimal-control methods~\cite{Sklarz2004DissipativeDynamics,PhysRevA.102.052605}. Effective non-Hermitian and open-system PMP formulations provide related directions~\cite{PhysRevA.107.022216,Lokutsievskiy_2024}. More flexible pulse parametrizations could be incorporated using GRAPE- or Krotov-type schemes~\cite{deFouquieres2011SecondOrderGRAPE,Khaneja2005GRAPE,Reich2012KrotovJCP}. Robustness criteria could also be included directly in the objective functional, following broader developments in quantum optimal control~\cite{Glaser2015TrainingSchrodingersCat,Mahesh2023QuantumOptimalControl,Ansel_2024}. These directions may also benefit from data-driven, model-free, and reinforcement-learning-based quantum-control strategies~\cite{Sivak2022PRXModelFreeQuantumControlRL,Sarma2025PRAppRLAnsatzeDressedLogicalQubits}.

Because the present approach is built on a generic multilevel chain model, it is expected to apply to other state-transfer settings where adiabatic passage, dark-state transport, and leakage suppression must be treated together. Representative applications include STIRAP in atoms and molecules, as demonstrated in early experiments and summarized in broader reviews of adiabatic passage~\cite{Kuklinski1989,Gaubatz1990,Bergmann1998,Vitanov2001,Bergmann2015,Vitanov2017,Kral2007}. The same idea is also relevant to multistate dark-state transport~\cite{PembertonRoss2010DarkStates,Kumar2012DarkStateManifold}, optimized adiabatic-passage protocols~\cite{PhysRevA.86.023406,PhysRevA.80.042325}, and STIRAP-type dynamics affected by fluctuations, dissipation, and dephasing~\cite{Yatsenko2014,Hou2013,Blekos2020}. More generally, the same strategy may be extended to dressed-state and counterdiabatic shortcuts~\cite{Baksic2016,Li2016}, experimental shortcut-to-adiabatic-passage protocols~\cite{Du2016}, and broader shortcut-to-adiabaticity frameworks~\cite{GueryOdelin2019,Evangelakos2023STIRAPShortcuts,Cepaite2023COLD}. Leakage-suppressed control in superconducting multilevel platforms is another natural direction. STIRAP and coherent population transfer have been demonstrated in superconducting circuits~\cite{Kumar2016,Xu2016}, and superadiabatic transfer has also been implemented~\cite{Vepsalainen2019}. Related optimal-control STIRAP-type protocols have been explored in superconducting qudits~\cite{Zheng2022npjQI_OCT_STIRAP_Qudit,Niu2022PRAppSTIRUP}, and STIRAP-inspired robust gates have been proposed for superconducting dual-rail qubits~\cite{Singhal2025PRAppSTIRAPInspiredGatesDualRail}. 
This connection also suggests that leakage-suppressed PMP optimization may be useful for STIRAP-based geometric and holonomic gates, where leakage out of the intended dark-state manifold can corrupt the acquired geometric phase or holonomy and reduce the gate fidelity~\cite{Moller2007GeometricPhaseSTIRAP,Setiawan2023GeometricSTIRAP}.
Extensions to qudit gate design and synthesis can also be envisaged~\cite{Omanakuttan2023QuditEntanglers,Seifert2023PRAFluxoniumQuquartOCT}. 
We therefore expect this framework to provide a useful starting point for bridging idealized adiabatic-control protocols and realistic multilevel quantum devices.

\section*{Acknowledgement}
The research is supported by the National Natural Science Foundation of China (No. 12574082, No. 12174379, No. E31Q02BG), the Chinese Academy of Sciences (No. E0SEBB11, No. E27RBB11), Quantum Science and Technology-National Science and Technology Major Project (No. 2021ZD0302300) and Chinese Academy of Sciences Project for Young Scientists in Basic Research (YSBR-090).

% ========= appendices =========
\appendix
\begin{widetext}
\section{First-Order Optimality Conditions for the Bolza Problem}
\label{app:kkt_pmp}

In this appendix, we derive a practical form of Pontryagin's maximum principle (PMP) from variational calculus and clarify why the gradient formula used in the main text defines a descent direction. We consider the Bolza functional
\begin{equation}
J[x,u]=\phi\!\bigl(x(T)\bigr)+\int_{0}^{T}L\!\bigl(x(t),u(t)\bigr)\,dt,
\label{eq:bolza_cost}
\end{equation}
subject to the dynamical constraint
\begin{equation}
c(x,u)\equiv \dot{x}(t)-f\!\bigl(x(t),u(t)\bigr)=0,
\qquad
x(0)=x_0.
\label{eq:dynamics_constraint}
\end{equation}
Because the dynamics constitute an infinite-dimensional equality constraint in time, it is natural to introduce a time-dependent Lagrange multiplier $\lambda(t)$ and define the augmented Lagrangian
\begin{equation}
\mathcal{L}[x,u,\lambda]
=
\phi\!\bigl(x(T)\bigr)
+
\int_{0}^{T}
\left[
L(x,u)+\lambda^{T}\bigl(\dot{x}-f(x,u)\bigr)
\right]dt.
\label{eq:augmented_lagrangian}
\end{equation}
The first variation with respect to $x$ is
\begin{equation}
\delta_x\mathcal{L}
=
\phi_x\!\bigl(x(T)\bigr)\delta x(T)
+
\int_{0}^{T}
\left[
L_x\,\delta x+\lambda^{T}\bigl(\delta\dot{x}-f_x\,\delta x\bigr)
\right]dt,
\label{eq:variation_x_1}
\end{equation}
where the subscripts denote partial derivatives evaluated along the trajectory $(x(t),u(t))$. Integrating the term containing $\delta\dot{x}$ by parts yields
\begin{equation}
\int_{0}^{T}\lambda^{T}\delta\dot{x}\,dt
=
\lambda^{T}\delta x\biggr|_{0}^{T}
-
\int_{0}^{T}\dot{\lambda}^{T}\delta x\,dt.
\label{eq:ibp_term}
\end{equation}
Substituting Eq.~(\ref{eq:ibp_term}) into Eq.~(\ref{eq:variation_x_1}), we obtain
\begin{equation}
\delta_x\mathcal{L}
=
\bigl[\phi_x\!\bigl(x(T)\bigr)+\lambda^{T}(T)\bigr]\delta x(T)
-\lambda^{T}(0)\delta x(0)
+
\int_{0}^{T}
\bigl(
L_x-\lambda^{T}f_x-\dot{\lambda}^{T}
\bigr)\delta x\,dt.
\label{eq:variation_x_2}
\end{equation}
Because the initial state is fixed, $\delta x(0)=0$, whereas $\delta x(t)$ is otherwise arbitrary on $(0,T)$. Stationarity therefore yields the costate equation
\begin{equation}
\dot{\lambda}^{T}(t)
=
L_x\!\bigl(x(t),u(t)\bigr)
-
\lambda^{T}(t)\,f_x\!\bigl(x(t),u(t)\bigr),
\label{eq:costate_equation}
\end{equation}
together with the terminal condition
\begin{equation}
\lambda^{T}(T)
=
-\phi_x\!\bigl(x(T)\bigr).
\label{eq:terminal_condition}
\end{equation}
We then define the Pontryagin Hamiltonian
\begin{equation}
\mathcal{H}_{\mathrm{P}}(x,u,\lambda)
\equiv h(x,u,\lambda)
=
\lambda^{T}f(x,u)-L(x,u).
\label{eq:pmp_hamiltonian}
\end{equation}
Once Eqs.~(\ref{eq:costate_equation}) and (\ref{eq:terminal_condition}) are imposed, the first variation with respect to the control reduces to
\begin{equation}
\delta\mathcal{L}
=
\int_{0}^{T}
\bigl(
L_u-\lambda^{T}f_u
\bigr)\delta u\,dt
=
-
\int_{0}^{T}
\frac{\partial h}{\partial u}\,\delta u\,dt.
\label{eq:variation_u}
\end{equation}
Hence, choosing the control variation as
\begin{equation}
\delta u(t)
=
\epsilon\,
\frac{\partial h}{\partial u}\bigl(x(t),u(t),\lambda(t)\bigr),
\qquad
\epsilon>0,
\label{eq:gradient_update}
\end{equation}
gives
\begin{equation}
\delta\mathcal{L}
=
-\epsilon
\int_{0}^{T}
\left\|
\frac{\partial h}{\partial u}
\right\|^{2}dt
\le 0.
\label{eq:descent_property}
\end{equation}
Therefore, the PMP gradient defines a local descent direction and drives the objective toward a stationary point of the constrained optimization problem.

\section{Trust-Region Method, Dogleg Strategy, and BFGS Quasi-Newton Update}
\label{app:trust_region_bfgs}

This appendix summarizes the trust-region framework, the BFGS quasi-Newton curvature update, and the Dogleg strategy used in our numerical optimization
\cite{Powell1970Dogleg,Fletcher1970BFGS,MoreSorensen1983TrustRegion,Yuan2015TrustRegionAdvances}.
Consider the unconstrained problem
\begin{equation}
\min_{x\in\mathbb{R}^{n}} f(x),
\label{eq:app_opt_problem}
\end{equation}
where $f(x)$ is continuously differentiable and twice differentiable when needed. At the $k$th iterate $x_k$, with gradient $g_k=\nabla f(x_k)$, we define the local quadratic model
\begin{equation}
m_k(p)
=
f(x_k)+g_k^{T}p+\frac{1}{2}p^{T}B_k p,
\label{eq:app_tr_model}
\end{equation}
where $p$ is the trial step and $B_k$ is the Hessian or a symmetric approximation to it. The trust-region subproblem is
\begin{equation}
\min_{p\in\mathbb{R}^{n}} m_k(p)
\qquad
\text{s.t.}
\qquad
\|p\|\le \Delta_k,
\label{eq:app_tr_subproblem}
\end{equation}
with trust-region radius $\Delta_k>0$. This constraint reflects the fact that the quadratic model is reliable only within a local neighborhood of $x_k$.

In practice, constructing the exact Hessian is often expensive, so we update $B_k$ using the BFGS formula. Defining
\begin{equation}
s_k=x_{k+1}-x_k,
\qquad
y_k=g_{k+1}-g_k,
\label{eq:app_bfgs_sy}
\end{equation}
the BFGS update is
\begin{equation}
B_{k+1}
=
B_k
-
\frac{B_k s_k s_k^{T} B_k}{s_k^{T}B_k s_k}
+
\frac{y_k y_k^{T}}{y_k^{T}s_k}.
\label{eq:app_bfgs_update}
\end{equation}
This update satisfies the secant condition
\begin{equation}
B_{k+1}s_k=y_k,
\label{eq:app_secant}
\end{equation}
and preserves positive definiteness provided that $B_0$ is symmetric positive definite and the curvature condition
\begin{equation}
y_k^{T}s_k>0
\label{eq:app_curvature_condition}
\end{equation}
holds. This property is particularly important for the Dogleg method, which relies on a positive-definite quadratic model.

Given a candidate step $p_k$, the trust-region method compares the actual decrease in $f$ with the decrease predicted by the model:
\begin{equation}
\rho_k
=
\frac{f(x_k)-f(x_k+p_k)}
{m_k(0)-m_k(p_k)}.
\label{eq:app_rho_def}
\end{equation}
A standard radius-update rule is
\begin{equation}
\Delta_{k+1}
=
\begin{cases}
\frac{1}{4}\Delta_k, & \rho_k<\frac{1}{4}, \\[6pt]
\min(2\Delta_k,\hat{\Delta}), & \rho_k>\frac{3}{4}\ \text{and}\ \|p_k\|=\Delta_k, \\[6pt]
\Delta_k, & \text{otherwise},
\end{cases}
\label{eq:app_radius_update}
\end{equation}
where $\hat{\Delta}$ is the prescribed maximum radius. The iterate is updated according to
\begin{equation}
x_{k+1}
=
\begin{cases}
x_k+p_k, & \rho_k>\eta, \\[6pt]
x_k, & \rho_k\le \eta,
\end{cases}
\label{eq:app_acceptance}
\end{equation}
with acceptance threshold $0<\eta<1$. In this way, the algorithm suppresses unreliable large steps far from the minimum while recovering fast local convergence near it.

To solve the trust-region subproblem efficiently, we use the Dogleg method. Let
\begin{equation}
p_k^{N}=-B_k^{-1}g_k
\label{eq:app_newton_step}
\end{equation}
denote the Newton step, which minimizes $m_k(p)$ in the absence of the trust-region constraint. If $\|p_k^{N}\|\le \Delta_k$, we simply take
\begin{equation}
p_k=p_k^{N}.
\label{eq:app_dogleg_case_newton}
\end{equation}
Along the steepest-descent direction $-g_k$, minimizing the quadratic model yields
\begin{equation}
p_k^{U}
=
-\frac{g_k^{T}g_k}{g_k^{T}B_k g_k}\,g_k,
\label{eq:app_dogleg_pu}
\end{equation}
which is the optimal one-dimensional step along $-g_k$. The Dogleg path is then defined by
\begin{equation}
\tilde{p}_k(\tau)
=
\begin{cases}
\tau p_k^{U}, & 0\le \tau \le 1, \\[6pt]
p_k^{U}+(\tau-1)(p_k^{N}-p_k^{U}), & 1\le \tau \le 2.
\end{cases}
\label{eq:app_dogleg_path}
\end{equation}

If $\|p_k^{U}\|\ge \Delta_k$, the trial step is taken on the boundary along the steepest-descent direction:
\begin{equation}
p_k
=
-\Delta_k\,\frac{g_k}{\|g_k\|}.
\label{eq:app_dogleg_case_gradient}
\end{equation}
If $\|p_k^{U}\|<\Delta_k<\|p_k^{N}\|$, the accepted step lies on the second segment of the Dogleg path,
\begin{equation}
p_k
=
p_k^{U}+\tau_k(p_k^{N}-p_k^{U}),
\qquad
0<\tau_k<1,
\label{eq:app_dogleg_boundary}
\end{equation}
where $\tau_k$ is determined by the boundary condition
\begin{equation}
\|p_k^{U}+\tau_k(p_k^{N}-p_k^{U})\|^{2}
=
\Delta_k^{2}.
\label{eq:app_dogleg_tau_eq}
\end{equation}
Defining $d_k=p_k^{N}-p_k^{U}$, this is equivalent to the quadratic equation
\begin{equation}
\|d_k\|^{2}\tau_k^{2}
+
2(p_k^{U})^{T}d_k\,\tau_k
+
\bigl(\|p_k^{U}\|^{2}-\Delta_k^{2}\bigr)
=
0,
\label{eq:app_dogleg_tau_quad}
\end{equation}
whose root in $(0,1)$ gives the required boundary intersection.

In summary, the trust-region framework specifies the local region in which the quadratic model is trusted and provides the step-acceptance criterion, BFGS supplies an efficient curvature approximation, and Dogleg solves the resulting subproblem at low cost while preserving favorable global and local convergence properties.

\section{Circuit Hamiltonian, Sixth-Order Approximation, and Level Spectrum}
The Xmon circuit is described by a pair of canonically conjugate operators: the superconducting phase difference $\hat{\phi}$ across the Josephson junction and the Cooper-pair number operator $\hat{n}$, which satisfy $[\hat{\phi},\hat{n}]=i$. In the absence of external flux bias, the bare circuit Hamiltonian is
\begin{equation}
H
=
4E_C\hat{n}^{\,2}-E_J\cos\hat{\phi},
\label{eq:HX}
\end{equation}
where $E_C$ and $E_J$ denote the charging and Josephson energies, respectively. In the transmon regime $E_J\gg E_C$, phase fluctuations are small, so the cosine potential may be expanded about $\hat{\phi}=0$ up to sixth order:
\begin{equation}
-E_J\cos\hat{\phi}
\approx
-E_J+\frac{E_J}{2}\hat{\phi}^{2}
-\frac{E_J}{24}\hat{\phi}^{4}
+\frac{E_J}{720}\hat{\phi}^{6}.
\label{eq:cos_expand_6}
\end{equation}
Discarding the constant offset $-E_J$, we obtain the effective sixth-order Hamiltonian
\begin{equation}
H^{(6)}
=
4E_C\hat{n}^{\,2}
+\frac{E_J}{2}\hat{\phi}^{2}
-\frac{E_J}{24}\hat{\phi}^{4}
+\frac{E_J}{720}\hat{\phi}^{6}.
\label{eq:H6}
\end{equation}

To diagonalize the quadratic part, we introduce bosonic ladder operators $a$ and $a^{\dagger}$ and define the zero-point fluctuation amplitudes
\begin{equation}
\phi_{\mathrm{zpf}}=\left(\frac{2E_C}{E_J}\right)^{1/4},
\qquad
n_{\mathrm{zpf}}=\left(\frac{E_J}{32E_C}\right)^{1/4}.
\label{eq:zpf_def}
\end{equation}
The canonical operators are then written as
\begin{equation}
\hat{\phi}
=
\phi_{\mathrm{zpf}}(a+a^{\dagger}),
\qquad
\hat{n}
=
-i\,n_{\mathrm{zpf}}(a-a^{\dagger}).
\label{eq:phin_zpf}
\end{equation}
Substituting these expressions into the quadratic Hamiltonian gives
\begin{equation}
H^{(2)}
=
4E_C\hat{n}^{\,2}
+\frac{E_J}{2}\hat{\phi}^{2}
=
\hbar\omega_0\left(a^{\dagger}a+\frac{1}{2}\right),
\label{eq:H2}
\end{equation}
with
\begin{equation}
\omega_0=\frac{\sqrt{8E_JE_C}}{\hbar}.
\label{eq:omega0_def}
\end{equation}
\label{app:transmon}
After including the quartic and sextic terms and retaining the weakly nonlinear contributions within the rotating-wave approximation, the Hamiltonian can be expressed as a polynomial in the number operator. Introducing the small parameter
\begin{equation}
\xi=\sqrt{\frac{2E_C}{E_J}},
\label{eq:xi_def}
\end{equation}
we obtain
\begin{equation}
\hat{H}^{(6)}
=
\left(
\hbar\omega_0-\frac{E_C}{2}+\frac{E_C\xi}{9}
\right)\hat{a}^{\dagger}\hat{a}
+
\left(
\frac{E_C\xi}{12}-\frac{E_C}{2}
\right)(\hat{a}^{\dagger}\hat{a})^{2}
+
\frac{E_C\xi}{18}(\hat{a}^{\dagger}\hat{a})^{3}.
\label{eq:H6_numberpoly}
\end{equation}
Accordingly, the energy of level $|n\rangle$ within the sixth-order approximation is
\begin{equation}
E_n
=
\langle n|\hat{H}^{(6)}|n\rangle
=
\left(
\hbar\omega_0-\frac{E_C}{2}+\frac{E_C\xi}{9}
\right)n
+
\left(
\frac{E_C\xi}{12}-\frac{E_C}{2}
\right)n^{2}
+
\frac{E_C\xi}{18}n^{3},
\label{eq:En_def}
\end{equation}
and the adjacent transition frequencies are
\begin{equation}
\omega_{n+1,n}
=
\frac{E_{n+1}-E_n}{\hbar},
\qquad
n=0,1,2,3.
\label{eq:omega_nn1}
\end{equation}
In the five-level truncation used in this work, the drift Hamiltonian is taken as Eq.~(\ref{eq:H0}).

In the eigenbasis $\{|n\rangle\}$, the charge operator in the transmon regime is dominated by nearest-neighbor matrix elements with harmonic-oscillator scaling,
\begin{equation}
\langle j-1|\hat{n}|j\rangle
\approx
i\,n_{\mathrm{zpf}}\sqrt{j},
\qquad
j=1,2,3,4.
\label{eq:n_matrix_element}
\end{equation}
Hence, within the truncated basis $\{|n\rangle\}_{n=0}^{4}$,
\begin{equation}
\hat{n}
\approx
i\,n_{\mathrm{zpf}}
\sum_{j=1}^{4}
\sqrt{j}
\Bigl(
|j-1\rangle\langle j|
-
|j\rangle\langle j-1|
\Bigr).
\label{eq:n_trunc}
\end{equation}
Combining Eq.~(\ref{eq:n_trunc}) with the drive term in Eq.~(\ref{eq:Hd}) immediately shows that all adjacent transitions are driven simultaneously.

Substituting the two-tone voltage drive of Eq.~(\ref{eq:V_two_tone}) into the control Hamiltonian and transforming to the interaction picture with respect to $H_0$, one obtains
\begin{equation}
H_I(t)
=
\frac{1}{2}
\sum_{j=1}^{4}
\sqrt{j}
\Bigl[
\Omega_p(t)e^{i\delta_p^{(j)}(t)}
+
\Omega_s(t)e^{i\delta_s^{(j)}(t)}
\Bigr]
|j-1\rangle\langle j|
+
\mathrm{h.c.},
\label{eq:HI_general}
\end{equation}
where $\Omega_p(t)$ and $\Omega_s(t)$ are the effective Rabi envelopes proportional to $2e\,n_{\mathrm{zpf}}V_p(t)$ and $2e\,n_{\mathrm{zpf}}V_s(t)$. Here,
$\mathrm{h.c.}$ denotes the Hermitian-conjugate term.
The detuning phases are
\begin{equation}
\delta_p^{(j)}(t)
=
\bigl(\omega_{j,j-1}-\omega_p\bigr)t+\phi_p,
\label{eq:delta_p_def}
\end{equation}
\begin{equation}
\delta_s^{(j)}(t)
=
\bigl(\omega_{j,j-1}-\omega_s\bigr)t+\phi_s,
\label{eq:delta_s_def}
\end{equation}
with $\omega_{j,j-1}=(E_j-E_{j-1})/\hbar$. This form makes explicit that the same two-tone drive not only addresses the target transitions $|0\rangle\leftrightarrow|1\rangle$ and $|1\rangle\leftrightarrow|2\rangle$, but also off-resonantly drives higher transitions and thereby opens leakage channels.

To obtain an approximately time-independent tridiagonal chain Hamiltonian, we introduce the diagonal unitary transformation
\begin{equation}
U(t)
=
\sum_{n=0}^{4}e^{-i\nu_n t}|n\rangle\langle n|,
\label{eq:U_rot}
\end{equation}
with cumulative reference frequencies
\begin{equation}
\nu_0=0,
\qquad
\nu_1=\omega_p,
\qquad
\nu_2=\omega_p+\omega_s,
\qquad
\nu_3=2\omega_p+\omega_s,
\qquad
\nu_4=2\omega_p+2\omega_s.
\label{eq:nu_list}
\end{equation}
The rotating-frame Hamiltonian is then
\begin{equation}
H_{\mathrm{rot}}(t)
=
U^{\dagger}(t)\,H(t)\,U(t)
-
i\hbar\,U^{\dagger}(t)\dot{U}(t),
\label{eq:Hrot_def}
\end{equation}
where $H(t)=H_0+H_c(t)$. Since $U(t)$ is diagonal, the diagonal terms define the effective detunings
\begin{equation}
\Delta_n
=
\frac{E_n}{\hbar}-\nu_n,
\qquad
n=0,1,2,3,4.
\label{eq:Delta_n_def}
\end{equation}
This yields the chain relations
\begin{equation}
\Delta_1
=
\omega_{10}-\omega_p,
\label{eq:Delta_1_chain}
\end{equation}
\begin{equation}
\Delta_2
=
\omega_{20}-(\omega_p+\omega_s)
=
\delta,
\label{eq:Delta_2_chain}
\end{equation}
\begin{equation}
\Delta_3
=
\omega_{30}-(2\omega_p+\omega_s),
\label{eq:Delta_3_chain}
\end{equation}
\begin{equation}
\Delta_4
=
\omega_{40}-(2\omega_p+2\omega_s).
\label{eq:Delta_4_chain}
\end{equation}
Under the rotating-wave approximation, one then recovers the five-level chain Hamiltonian of Eq.~(\ref{eq:H_chain_main}), whose couplings inherit the $\sqrt{j}$ scaling from the charge-operator matrix elements.

\section{Computational complexity of the PMP optimization}
\label{App:complexity}

This appendix estimates the computational cost of the PMP-based optimization used in this work. The purpose is to clarify how the cost scales with the number of retained levels, time steps, control parameters, and optimization iterations. It also explains why we optimize a low-dimensional set of Gaussian pulse parameters rather than a high-dimensional time-sliced waveform.

Let $N$ be the number of retained levels in the truncated multilevel chain, $M$ the number of time steps, $p$ the number of control parameters, and $K$ the number of outer optimization iterations. In this work, the two control pulses are parametrized by six Gaussian parameters,
\begin{equation}
\mathbf{u}
=
(A_p,A_s,t_{0p},t_{0s},\sigma_p,\sigma_s),
\qquad p=6.
\end{equation}
For each PMP gradient evaluation, one forward propagation of the state $|\psi(t)\rangle$ and one backward propagation of the costate $\langle\lambda(t)|$ are required. The parameter gradients are accumulated on the same time grid. If $C_H(N)$ denotes the cost of applying the Hamiltonian to a state vector, then
\begin{equation}
C_{\mathrm{grad}}
=
O\!\left(2M C_H(N)+Mp\right).
\end{equation}
The term $Mp$ comes from accumulating the gradients of the $p$ pulse parameters. For Gaussian pulses, these derivatives are analytic, and the corresponding overhead is small.

The parameter update is performed using a trust-region method with a BFGS approximation to the Hessian. Storing the BFGS matrix requires $O(p^2)$ memory, while directly solving the associated dogleg or Newton-type subproblem requires $O(p^3)$ operations. If each outer iteration requires, on average, $r$ trial objective evaluations, with each trial requiring one additional forward propagation, the cost per iteration is
\begin{equation}
C_{\mathrm{iter}}
=
O\!\left[(2+r)M C_H(N)+Mp+p^3\right].
\end{equation}
Thus, the total cost scales as
\begin{equation}
C_{\mathrm{tot}}
=
O\!\left\{
K\left[(2+r)M C_H(N)+Mp+p^3\right]
\right\}.
\end{equation}
The desired numerical precision mainly affects the time resolution $M$, the propagation tolerance, and the number of optimization iterations $K$.

For the nearest-neighbor chain Hamiltonian used here, the Hamiltonian is sparse and tridiagonal, so one Hamiltonian-vector multiplication scales as $C_H(N)=O(N)$. Therefore,
\begin{equation}
C_{\mathrm{iter}}
=
O\!\left[(2+r)MN+Mp+p^3\right].
\end{equation}
If a dense Hamiltonian representation is used instead, $C_H(N)=O(N^2)$, giving
\begin{equation}
C_{\mathrm{iter}}
=
O\!\left[(2+r)MN^2+Mp+p^3\right].
\end{equation}
When the forward trajectory is stored for the backward costate propagation, the memory cost is
\begin{equation}
S_{\mathrm{tot}}
=
O(MN+p^2).
\end{equation}

For the five-level transmon example considered in this work, $N=5$ and $p=6$. If the trust-region step requires about one trial objective evaluation per iteration, namely $r\simeq 1$, the sparse-chain estimate reduces to
\begin{equation}
C_{\mathrm{iter}}
=
O\!\left(3MN+Mp+p^3\right).
\end{equation}
Because $p=6$ is small, the cost of the trust-region, BFGS, and Dogleg subproblems is negligible compared with the state and costate propagations. The present PMP optimization is therefore numerically lightweight for the system sizes studied here.

It is useful to contrast this with a free time-sliced control parametrization. If the control amplitude at each time step is treated as an independent variable, then for $q$ control channels the number of parameters becomes
\begin{equation}
p_{\mathrm{free}}\sim qM .
\end{equation}
Such an approach is possible in principle and is closely related to a discretized weak-PMP condition or to GRAPE-type updates. However, it greatly increases the cost of parameter-space operations. For example, full BFGS storage would scale as $O(q^2M^2)$, and a direct Dogleg or Newton solve would scale as $O(q^3M^3)$. The Gaussian parametrization used in this work avoids this overhead. As a result, the optimization cost is mainly set by state and costate propagation, rather than by large matrix operations in the control-parameter space. Therefore, the low-dimensional Gaussian ansatz is not a limitation of the PMP formalism itself, but a deliberate choice made for the present experimentally constrained STIRAP optimization.

\section{Adiabaticity analysis of the optimized protocol}
\label{App:adiabaticity}

This appendix compares the adiabaticity and leakage behavior of the initial STIRAP protocol and the PMP-optimized protocol. The purpose is not to show that the optimized protocol reaches the strict adiabatic limit. Instead, we distinguish the ideal three-level STIRAP adiabaticity from the dressed-state following in the full five-level transmon system. This comparison shows that a pulse sequence that is adiabatic in the ideal three-level subspace does not necessarily guarantee low leakage in the full multilevel dynamics.

In the ideal three-level STIRAP model, the instantaneous dark state in the target subspace $\{\ket{0},\ket{1},\ket{2}\}$ is
\begin{equation}
\ket{D_{\mathrm{tar}}(t)}
=
\frac{
\Omega_{12}(t)\ket{0}-\Omega_{01}(t)\ket{2}
}{
\sqrt{|\Omega_{01}(t)|^2+|\Omega_{12}(t)|^2}
}.
\label{eq:app_Dtar}
\end{equation}
The corresponding three-level adiabatic parameter is
\begin{equation}
\mathcal{A}_{\theta}(t)
=
\frac{|\dot{\theta}(t)|}{\Omega_{\mathrm{eff}}(t)},
\qquad
\tan\theta(t)=
\frac{|\Omega_{01}(t)|}{|\Omega_{12}(t)|},
\qquad
\Omega_{\mathrm{eff}}(t)
=
\sqrt{|\Omega_{01}(t)|^2+|\Omega_{12}(t)|^2}.
\label{eq:app_Atheta}
\end{equation}
This quantity measures how slowly the mixing angle of the ideal dark state changes. However, it does not include leakage couplings, dressed-state gap variations, or multilevel nonadiabatic couplings. Therefore, $\mathcal{A}_{\theta}$ should only be viewed as a diagnostic within the ideal three-level STIRAP subspace.

To characterize dark-state following in the full multilevel system, we diagonalize the full coherent Hamiltonian $\tilde{H}(t)$ at each time. The dressed dark state is chosen as the instantaneous eigenstate with the largest overlap with the ideal three-level dark state,
\begin{equation}
\ket{\phi_D(t)}
=
\ket{\phi_{\ell_D}(t)},
\qquad
\ell_D(t)
=
\operatorname*{arg\,max}_{\ell}
\left|
\braket{\phi_{\ell}(t)}{D_{\mathrm{tar}}(t)}
\right|^2 .
\label{eq:app_dressed_dark}
\end{equation}
The overlap between the evolving state and this dressed dark-state channel is measured by
\begin{equation}
P_D^{\mathrm{full}}(t)
=
\left|
\left\langle
\phi_D(t)
\middle|
\frac{\psi(t)}{\sqrt{\braket{\psi(t)}{\psi(t)}}}
\right\rangle
\right|^2 .
\label{eq:app_PDfull}
\end{equation}
We further define the dressed-state adiabatic parameter of the full five-level system as
\begin{equation}
\mathcal{A}_{\mathrm{full}}(t)
=
\max_{\ell\neq D}
\frac{
\left|
\mel{\phi_{\ell}(t)}{\partial_t}{\phi_D(t)}
\right|
}{
|E_{\ell}(t)-E_D(t)|/\hbar
}.
\label{eq:app_Afull}
\end{equation}
This quantity directly measures the nonadiabatic coupling between the dressed dark-state channel and the other instantaneous eigenstates of the full Hamiltonian.

Table~\ref{tab:app_adiabatic_summary} summarizes the numerical diagnostics. The initial STIRAP protocol has a clear dark-state transfer structure in the ideal three-level subspace, with $\max\mathcal{A}_{\theta}=0.3248$. However, in the full five-level system, the same pulses do not lead to strongly adiabatic dressed-state evolution. The full dressed-state parameter reaches $\max\mathcal{A}_{\mathrm{full}}=2.424$, and the maximum leakage population is $\max_t P_{\mathrm{leak}}(t)=0.04986$. This shows that three-level STIRAP adiabaticity alone is not sufficient to ensure low-leakage evolution in the full transmon system.

\begin{table}[t]
\centering
\caption{
Adiabaticity and leakage diagnostics for the initial and PMP-optimized protocols.
Here $\mathcal{A}_{\theta}$ is the adiabatic parameter in the ideal three-level STIRAP subspace, while $\mathcal{A}_{\mathrm{full}}$ is the dressed-state adiabatic parameter in the full five-level system.
}
\label{tab:app_adiabatic_summary}
\begin{tabular}{c|c|c}
\hline
Quantity & Initial protocol & PMP-optimized protocol \\
\hline
$F$ & $0.9113$ & $0.9982$ \\
$1-F$ & $0.08866$ & $1.84\times 10^{-3}$ \\
$T~(\mathrm{ns})$ & $80$ & $47.78$ \\
$\max_t P_{\mathrm{leak}}(t)$ & $0.04986$ & $0.02211$ \\
$\overline{P_D^{\mathrm{full}}}$ & $0.8951$ & $0.9172$ \\
$\max \mathcal{A}_{\theta}$ & $0.3248$ & $0.7176$ \\
$\max \mathcal{A}_{\mathrm{full}}$ & $2.424$ & $1.025$ \\
\hline
\end{tabular}
\end{table}

After PMP optimization, the protocol duration is reduced from $80~\mathrm{ns}$ to $47.78~\mathrm{ns}$. Because the transfer is faster, the three-level mixing angle changes more rapidly, and $\max\mathcal{A}_{\theta}$ increases from $0.3248$ to $0.7176$. Thus, the optimized protocol should not be interpreted as a simple improvement of the ideal three-level adiabaticity. The main improvement occurs in the full dressed-state description. The value of $\max\mathcal{A}_{\mathrm{full}}$ decreases from $2.424$ to $1.025$, and the average dressed dark-state population increases from $0.8951$ to $0.9172$. At the same time, the maximum leakage population decreases from $0.04986$ to $0.02211$, and the final fidelity increases from $0.9113$ to $0.9982$.

These results indicate that the PMP-optimized protocol preserves a STIRAP-like dark-state transfer structure. Its main role, however, is not to make the ideal three-level STIRAP process more adiabatic. Instead, it improves dressed-state following and suppresses leakage in the full five-level system within a shorter time. The optimized protocol should therefore be understood as a finite-time, leakage-suppressed STIRAP-like protocol, rather than as a slow STIRAP protocol in the strict adiabatic limit.
\end{widetext}

% ========= bibliography =========
\newpage
\bibliography{ref}

\end{document}